\documentclass[prb,aps,twocolumn,showpacs,superscriptaddress,amsmath,amssymb]{revtex4}

\usepackage{graphicx}
\usepackage{psfrag}
\usepackage{epsfig}
\usepackage{color}
\usepackage{subfigure}
\definecolor{dred}{rgb}{0.7,0.0,0.0}
\usepackage{dcolumn}% Align table columns on decimal point
\usepackage{bm}% bold math

\begin{document}

\title {      $t$--$J$ model of coupled Cu$_2$O$_5$ ladders
              in Sr$_{14-x}$Ca$_x$Cu$_{24}$O$_{41}$ }

\author {     Krzysztof Wohlfeld }
\affiliation{ IFW Dresden, P. O. Box 270116, D-01171 Dresden, Germany }
\affiliation{ Marian Smoluchowski Institute of Physics, Jagellonian
              University, Reymonta 4, PL-30059 Krak\'ow, Poland }

\author {     Andrzej M. Ole\'{s} }
\affiliation{ Marian Smoluchowski Institute of Physics,
              Jagellonian University, Reymonta 4,
              PL-30059 Krak\'ow, Poland }
\affiliation{ Max-Planck-Institut f\"ur Festk\"orperforschung,
              Heisenbergstrasse 1, D-70569 Stuttgart, Germany }

\author {     George A. Sawatzky }
\affiliation{ Department of Physics and Astronomy, University of
              British Columbia, Vancouver B. C. V6T-1Z1, Canada}

\date{\today}

\begin{abstract}

Starting from the proper charge transfer model for Cu$_2$O$_5$ coupled
ladders in Sr$_{14-x}$Ca$_x$Cu$_{24}$O$_{41}$ we derive the low energy
Hamiltonian for this system. It occurs that the widely used ladder
$t$--$J$ model is not sufficient and has to be supplemented by
the Coulomb repulsion term between holes in the neighboring ladders.
Furthermore, we show how a simple mean-field solution of the derived
$t$--$J$ model may explain the onset of the charge density wave with
the odd period in Sr$_{14-x}$Ca$_x$Cu$_{24}$O$_{41}$. 

{\it Published in Phys. Rev. B {\bf 81}, 214522 (2010)} 
\end{abstract}


\maketitle

\section{Introduction}
\label{sec:1}

It is widely assumed that the two-dimensional (2D) $t$--$J$ model
\cite{Cha78} is {\it the} correct model to describe the low-energy
physics of the CuO$_2$ planes.\cite{Zha88,Dag94} Consequently,
many authors believe that the high-temperature superconductivity
in the cuprates {\it can be} explained by this model and it is
merely the computationally challenging character of the model
which leads to the lack of the understanding of the
superconducting ground state (see e.g. Refs. \onlinecite{Mai05}).

Similarly, it has been suggested that the $t$--$J$ model defined
on the ladder (called ladder $t$--$J$ model in what follows)
is the right model to describe the low energy
physics relevant for the Cu$_2$O$_5$ coupled ladder planes of
Sr$_{14-x}$Ca$_x$Cu$_{24}$O$_{41}$ (SCCO).\cite{Dag96,Rus06} This
is a very attractive theoretical idea because: (i) the ladder
$t$--$J$ model is much easier to solve than its 2D
counterpart and it has a superconducting ground state for some
specific range of parameters,\cite{Dag92} (ii) a superconducting
ground state (under pressure of 3 GPa) was found\cite{Ueh96} in
the ladder planes of SCCO for $x=13.6$. This may suggest that
indeed the $t$--$J$ model contains the essential physics needed to
explain the superconductivity, at least in the ladders.

In this paper we would like to question the above point of view.
As we show below, the ladder $t$--$J$ model is too oversimplified
and thus not sufficient to describe the low energy physics of the
ladder planes in SCCO. Actually, this can already be inferred by
comparing the experimental observations in SCCO with the
theoretical predictions for the ladders:\cite{Woh07} (i) a charge
density wave (CDW) ground state with period $3$ and $5$ was
observed, but (ii) no CDW state with even period has been
found,\cite{Rus06} whereas (iii) the ladder $t$--$J$ model
may have a CDW ground state only with an even
period.\cite{Whi02,Rou07} Therefore, we investigate this problem
here by a systematic derivation of the proper $t$--$J$ model
for the coupled ladder system, extended for topological reasons by
the interladder repulsive term, and discuss a simple solution of
this model.

The paper is organized as follows. We introduce the charge
transfer Hamiltonian in Sec. \ref{sec:2}. Next, in Sec.
\ref{sec:3} we derive the low energy $t$--$J$
Hamiltonian which contains the kinetic energy and the superexchange,
similar to the ladder $t$--$J$ model, and the intraladder and
interladder repulsion terms. In Sec. \ref{sec:4} we examine the
role of the Coulomb intersite repulsion. Finally, we present a
numerical solution of the model in Sec. \ref{sec:5} and draw
conclusions in Sec. \ref{sec:6}. The paper is supplemented by two
appendices where some of the mathematical details of the model
derivation are discussed.

\section{The charge transfer Hamiltonian}
\label{sec:2}

As the starting point we choose the multiband charge transfer
Hamiltonian introduced before for the Cu$_2$O$_5$ coupled ladder
geometry in SCCO.\cite{Woh07} The model in hole notation reads,
\begin{equation}
\mathcal{H}= \mathcal{H}_0 + \mathcal{H}_1 + \mathcal{H}_2\,,
\label{eq:ct} \\
\end{equation}
\begin{eqnarray}
\mathcal{H}_0&\!=\!&-t_{pd} \sum_{i \alpha  \sigma}\!
\Big( d^\dag_{i \alpha \sigma}y_{i \alpha \sigma}^{} -d^\dag_{i+1, \alpha
\sigma}y_{i \alpha \sigma}^{}
 \mp  d^\dag_{i \alpha  \sigma}x_{i \alpha  \sigma}^{} \nonumber \\
& \pm& d^\dag_{i \alpha\sigma}b^{}_{i \sigma} +\mbox{H.c.} \Big)
+ \Delta\sum_{i \alpha} \Big( n_{i \alpha x} + n_{i \alpha y} \Big)
\nonumber \\
&+&\Delta\sum_{i} n_{ib}
+U\sum_{i \alpha} n_{i \alpha \uparrow} n_{i \alpha \downarrow}\,, \\
\mathcal{H}_1&\!=\!&U_p\sum_{i \alpha,\xi=x,y}
n_{i \alpha \xi\uparrow}n_{i \alpha \xi\downarrow}
+U_p\sum_{i} n_{ib\uparrow}n_{ib\downarrow}\,, \\
\mathcal{H}_2&\!=\!&U_p (1-2\eta)\sum_{i \alpha \sigma}
\!\Big( n_{i \alpha x\sigma}\bar{n}_{i \bar{\alpha} y\bar{\sigma}}
+n_{i \alpha y\sigma}\bar{n}_{i \bar{\alpha} x\bar{\sigma}} \Big)
\nonumber \\
&+&U_p(1-3\eta)\sum_{i \alpha \sigma} \Big(n_{i \alpha x\sigma}\bar{n}_{i
\bar{\alpha} y\sigma} \!+\!n_{i \alpha y\sigma}\bar{n}_{i \bar{\alpha} x\sigma}
\Big)\,.
\end{eqnarray}
The model (1) was adopted to the present ladder geometry\cite{Woh07}
from the charge transfer models introduced before for CuO$_2$
planes,\cite{Ole87} and CuO$_3$ chains\cite{Ole91} in high
temperature superconductors. The parameters are: the energy for
oxygen $2p_{\sigma}$ ($2p_x$ or $2p_y$ with creation operators
$\{x_{i\alpha\sigma}^{\dagger}\}$ and
$\{y_{i\alpha\sigma}^\dagger{}\}$) orbital $\Delta$ (the so-called
charge transfer energy measured with respect to the energy of $3d$
copper orbitals), the $d$-$p$ hopping $t_{pd}$ between the nearest
neighbor copper and oxygen sites, the on-site Coulomb repulsion
$U$ ($U_p$) on the copper (oxygen) sites, and $\eta=J_H/U_p \simeq
0.2$ --- a realistic value of Hund's exchange on oxygen
ions\cite{Gra92} (for a complete set of realistic parameters see
Sec. \ref{sec:para}). Besides, in principle the actual electron
energy at bridge orbital (rung position) of the ladder with
creation operators $\{b_{i\sigma}^{\dagger}\}$ is approximately 10
\% smaller than the one at other oxygen positions.\cite{Mue98}
However, it was shown\cite{Woh07b} that this difference does not
have any important physical consequences and therefore we will
neglect it here. Note also that the phases of the $\{3d,2p\}$
orbitals were explicitly taken into account in the hopping
elements $\propto t_{pd}$ (for clarity the phases are shown only
in Fig. \ref{fig:1}), the index $\alpha\in\{ R,L \}$ denotes the
right or left leg of the ladder ($\bar{R}=L$ and $\bar{L}=R$),
$\bar{\sigma}=-\sigma$ for $\sigma\in\{\uparrow,\downarrow\}$, and
the upper (lower) sign stands for terms with $\alpha=L$
($\alpha=R$).

%%%%%%%%%%%%%%%%%%%%%%%%%%%%%%%%%%%%%%%%%%%%%%%%%%%%%%%%
%%                       figure 0
%%%%%%%%%%%%%%%%%%%%%%%%%%%%%%%%%%%%%%%%%%%%%%%%%%%%%%%%
\begin{figure}[t!]
\includegraphics[width=8.4cm]{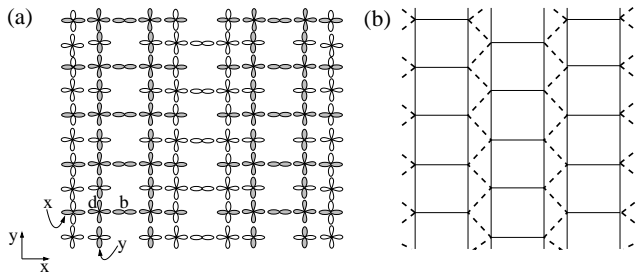}
\caption{
Schematic view of the coupled Cu$_2$O$_5$ (white/gray)
ladders in SCCO: (a) orbitals in charge transfer model (\ref{eq:ct});
(b) intraladder (interladder) bonds in the effective extended $t$--$J$
model, see Eq. (\ref{eq:tj}), shown by solid (dashed) lines.
}
\label{fig:0}
\end{figure}

The charge transfer model (1) includes seven orbitals per
Cu$_2$O$_5$ ladder unit cell $i$
(see Fig. \ref{fig:0}):
two Cu($3d_{x^2-y^2}\equiv d$) orbitals on the $R/L$
leg, two O($2p_y\equiv y$) orbitals on the $R/L$ leg,
two O($2p_x\equiv x$) side orbitals on the $R/L$ leg, and
one O($2p_x\equiv b$) bridge orbital on the rung.
Although it seems that the model is quasi one-dimensional (1D),
the density operators $\bar{n}_{i \alpha x \sigma}$
and $\bar{n}_{i \alpha y \sigma}$
stand for the oxygen hole densites in
the neighboring ladders and make it implicitly 2D as the
interladder coupling couples the ladders, so the model extends
over the entire Cu$_2$O$_5$ plane.

\section{The effective $t$--$J$ Hamiltonian}
\label{sec:3}

\subsection{The model and the superexchange}

In what follows we will derive the low-energy version of
Hamiltonian (\ref{eq:ct}) which is valid in the so-called charge
transfer regime $U>\Delta$, i.e. for the typical values of model
(\ref{eq:ct}) parameters: $U\simeq 8t_{pd}$, $\Delta\simeq
3t_{pd}$, and $U_p\simeq 3t_{pd}$, see Ref. \onlinecite{Hyb89} and
Sec. \ref{sec:para} below. The effective $t$--$J$ Hamiltonian
consists then of four terms, and may be thus also called
$t$--$J$--$V$ model,
\begin{align}\label{eq:tj}
H= H_t + H_J + H_{V_1}+H_{V_2}\,,
\end{align}
which are separately derived and discussed below.

We begin with the superexchange term $H_J$ which is the only
important term in Eq. (\ref{eq:tj}) at half-filling. In this case
(i.e. with one hole per copper site) the charge
transfer model (\ref{eq:ct}) can be easily reduced to the low
energy Heisenberg model for spins $S=1/2$ using the perturbation
theory to fourth order in $t_{pd}$:\cite{Zaa88}
\begin{align}
\label{eq:hj}
H_J&=J \sum_{i \alpha} \left({\bf S}_{i \alpha} \cdot {\bf S}_{i+1, \alpha} -
\frac{1}{4}\,\tilde{n}_{i\alpha} \tilde{n}_{i+1,\alpha} \right) \nonumber \\
&+J\sum_{ i} \left({\bf S}_{iR} \cdot {\bf S}_{iL} - \frac{1}{4}\,\tilde{n}_{iR}
\tilde{n}_{iL} \right)\,.
\end{align}
Here, tilde in $\tilde{n}_{i\alpha}$ implies that the hole double
occupancies are excluded. The superexchange constant contains
contributions due to charge excitations on copper sites and on the
intermediate oxygen site for a Cu--O--Cu bond, and for finite
$U_p$ reads:\cite{Zaa88}
\begin{equation}
J=\left(\frac{2t^2_{pd}}{\Delta}\right)^2
\left\{ \frac{1}{U} +\frac{2}{2\Delta+U_p} \right\}.
\end{equation}

One may wonder whether the geometry of coupled ladders could
influence the above result. Indeed, there exists a $90^\circ$
superexchange process between the holes on two neighboring
ladders. However, according to the Goodenough-Kanamori
rules\cite{Kan59} such a superexchange process is much weaker than
the superexchange generated by charge excitations along the
$180^\circ$ path in the single ladder and can be neglected.

\subsection{Zhang-Rice singlets for the ladder}

\begin{table}[b!]
\caption{Binding energy of the singlet and triplet state formed by
the copper hole and a doped hole in one of the three various
oxygen states: (i) symmetric plaquette state $|P_{i \alpha
\sigma}\rangle$, (ii) antisymmetric plaquette state $|A_{i \alpha
\sigma}\rangle$ orthogonal to Eq. (\ref{eq:pi}), and (iii) single
oxygen orbital. Here $t_1=t_{pd}^2/\Delta \sim t_{pd}/3$,
$t_2=t_{pd}^2/(U-\Delta) \sim t_{pd}/5$, and $t_3=t_1 U_p /
(\Delta+U_p) \sim t_{pd} /6$ for the typical charge transfer
parameters.\cite{Hyb89}} \label{tab:1}
\begin{ruledtabular}
\begin{tabular}{c c c c}
 &$|P_{i\alpha\sigma}\rangle$ & $|A_{i\alpha\sigma}\rangle$ &single oxygen\\ \hline
singlet  & $-8(t_1+t_2)+2t_3$ & $-4t_1+2t_3$  & $-2(t_1+t_2)+2t_3$
\cr triplet  & $0$                & $-4t_1$       & $0$
\cr
\end{tabular}
\end{ruledtabular}
\end{table}

When the ladder is away from half-filling, the perturbation theory
gets complicated. Therefore, following the Zhang and Rice
construction,\cite{Zha88} we first define a phase coherent {\it
symmetric plaquette state}\cite{symm} $|P_{i \alpha
\sigma}\rangle$ which is formed by the four oxygen orbitals
surrounding the central copper site $i \alpha$:
\begin{equation}
\label{eq:pi}
|P_{i \alpha \sigma}\rangle=
\frac{1}{2}\left(\pm x^\dag_{i \alpha \sigma} \mp b^\dag_{i \sigma}-
y^\dag_{i-1, \alpha \sigma}+y^\dag_{i \alpha \sigma}\right)|0\rangle\,,
\end{equation}
where the upper (lower) sign stands for $\alpha=L$ ($\alpha=R$).
When a hole in this state forms a singlet state with the hole at
the central copper site, it has a large negative binding energy of
$-8(t_1+t_2)+2t_3$, see caption of Table I for definition of $t_n$
hoppings and for more details. Actually, this binding energy is
not only much larger than the individual effective hopping terms
(which is of the order of $t_1$ or $t_2$, see Ref.
\onlinecite{Zha88}) but it is also considerably larger than the
binding energy of some other possible bound states, see Table I.
Note that finite $U_p$, not considered by Zhang and
Rice,\cite{Zha88} results in finite $t_3$ hopping but does not
qualitatively change the large binding energy of a symmetric
singlet state (\ref{eq:pi}).

The mere problem with the states defined by Eq. (\ref{eq:pi}) is
that they are not orthogonal. It can be checked that in the case
of the ladder geometry the following superposition of the
symmetric plaquette states forms a complete and orthogonal basis
for a low-energy Hilbert subspace:
\begin{equation} \label{eq:plaql}
\phi_{l\alpha \sigma}^{\dagger}|0\rangle
= \frac{1}{N}\sum_{jk} e^{ikl}e^{-ikj}
\left(\alpha_k |P_{j \alpha \sigma}\rangle
      +\beta_k |P_{j \bar{\alpha}\sigma}\rangle\right)\,,
\end{equation}
where $\alpha_k(\beta_k) = 2/\sqrt{3-2\cos k}\pm 2/\sqrt{5-2\cos k}$.
Then the Zhang-Rice (ZR) singlets for the ladder are:
\begin{equation} \label{eq:ZhangRice}
|\psi_{i\alpha}\rangle= \frac{1}{\sqrt{2}}\;
\left(\phi_{i\alpha\uparrow}^{\dagger}d_{i\alpha\downarrow}^{\dagger} -
\phi_{i\alpha\downarrow}^{\dagger}d_{i\alpha\uparrow}^{\dagger}\right)|0\rangle\,.
\end{equation}
Although the binding energy is slightly reduced after this
orthogonalization, the change is not significant:
If the energy splitting between the orthogonalized ZR singlets
and triplets is defined as $16\chi^2 t_1$
(we consider a simplified case $t_1=t_2$ and
$U_p=0$), then $\chi\approx 1$ --- both in the 1D
($\chi=0.98$) and in the 2D case ($\chi=0.96$),
see Ref. \onlinecite{Zha88}.

Having shown that the ZR singlets in the single ladder do not
differ much from those which arise in the 2D cuprates,\cite{Arr09}
we can now safely apply all the arguments used in Ref.
\onlinecite{Zha88} to derive the effective hopping of ZR singlets
following from finite $t_{pd}$. Thus, we obtain,
\begin{align}\label{eq:ht}
H_t&=-t \sum_{i \alpha \sigma } \Big\{ \tilde{d}^{\dag}_{i \alpha \sigma}
\tilde{d}^{}_{i \bar{\alpha} \sigma}+
\Big(\tilde{d}^{\dag}_{i \alpha \sigma} \tilde{d}^{}_{i+1, \alpha \sigma}
+ {\rm H.c.}\Big) \Big\}\,,
\end{align}
where once again $\tilde{d}_{i \alpha \sigma} = {d}_{i \alpha \sigma}
(1-n_{i\alpha \bar{\sigma}})$ is a fermion operator in the restricted
space.
While we do not show here the detailed expression for the effective
hopping $t$ of ZR singlets, note that it is considerably smaller than
$t_{pd}$ (ca. 50\%).\cite{Zha88} Note also that having two ZR singlets
at the same site costs energy $4t_2+2t_1$ (if $t_3=0$, see Ref.
\onlinecite{Zha88}) and therefore we used the tilde operators above
to exclude these local configurations of two ZR singlets.

\subsection{Intraladder repulsion $H_{V_1}$}

\begin{figure}[t!]
\includegraphics[width=0.3\textwidth]{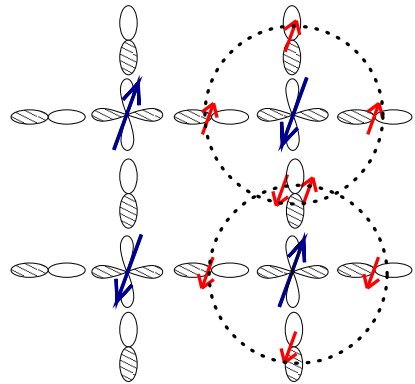}
\caption{(Color online) The artist's view of the intraladder
repulsion between two neighboring ZR singlets. Large (small)
arrows depict the hole spins for +1.0 (+0.25) charge --- they
stand for spins at copper sites and for the spins of doped holes
delocalized over oxygen orbitals. Orbital phases are depicted by
striped/white areas. } \label{fig:1}
\end{figure}

\begin{figure}[t!]
\includegraphics[width=0.47\textwidth]{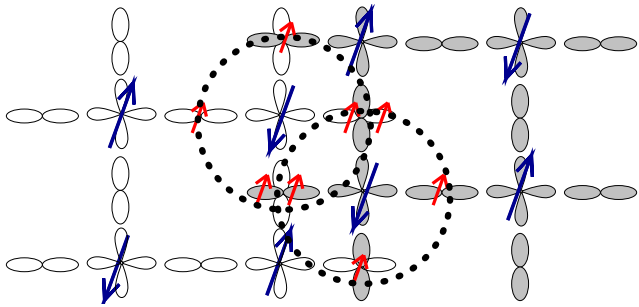}
\caption{(Color online) The artist's view of the interladder
repulsion between two ZR singlets on two different
(white and gray) ladders.
Spins are depicted similarly as in Fig. \ref{fig:1}.
} \label{fig:2}
\end{figure}

The Coulomb interaction on oxygen sites $U_p$, neglected in Ref.
\onlinecite{Zha88}, plays a minor role in the stability of the ZR
singlets (see e.g. finite $t_3$ for finite $U_p$ in Table I), but
this issue is more subtle.\cite{Fei95} Actually, due to finite
$U_p$ the two neighboring nonorthogonal ZR singlets repel each
other when two holes occupy a common oxygen site (see Figs.
\ref{fig:1} and \ref{fig:2}). Whereas the significance of the
interladder repulsion is discussed in the next subsection, let us
concentrate first on the repulsion between the ZR singlets {\it
within a single ladder} (see Fig. \ref{fig:1}), and calculate
repulsion $\propto U_p$ between two orthogonalized ZR singlets
within the ladder $\langle \psi_{s\alpha}, \psi_{r\beta} |
\mathcal{H}_1 | \psi_{h\beta}, \psi_{j\alpha} \rangle$. Let us
note that the `mixed terms' such as $\langle \psi_{s R},\psi_{r
L}|\mathcal{H}_1|\psi_{h L},\psi_{j L}\rangle$, which could {\it a
priori\/} destroy the ZR singlets, fortunately turn out to be much
smaller than the respective binding energy. Using Eq.
(\ref{eq:ZhangRice}), after a somewhat lengthy but straightforward
calculation (for more details see Appendix \ref{app:a}), one finds
the following values for the intraladder interaction along the leg
and the rung:
\begin{eqnarray}
 \langle \psi_{j\alpha}, \psi_{j+1,\alpha} | \mathcal{H}_1 |
\psi_{j+1,\alpha}, \psi_{j\alpha } \rangle&=&0.027\,U_p\,, \\
 \langle \psi_{j\alpha}, \psi_{j \bar{\alpha}} | \mathcal{H}_1 |
\psi_{j\bar{\alpha}}, \psi_{j\alpha } \rangle&=&0.026\,U_p\,.
\end{eqnarray}
We have verified that the interaction between the second nearest
neighbors is ca. 15 times smaller and can be safely neglected (the
longer-range interaction is even smaller, cf. Appendix
\ref{app:a}).

%%%%%%%%%%%%%%%%%%%%%%%%%%%%%%%%%%%%%%%%%%%%%%%%%%%%%%%%
%%                       figure 1
%%%%%%%%%%%%%%%%%%%%%%%%%%%%%%%%%%%%%%%%%%%%%%%%%%%%%%%%

Hence, one finds that the interaction among the nearest
neighbor ZR singlets is almost isotropic. Thus, we
can write the effective Hamiltonian for the repulsion
between ZR singlets
(as shown in Fig. \ref{fig:1})
\begin{align}\label{eq:hv1}
H_{V_1}=  V_1 \Big( \sum_{i\alpha} \tilde{n}_{i\alpha}
\tilde{n}_{i+1,\alpha} + \sum_{i}\tilde{n}_{i R}\tilde{n}_{i L} \Big)\,,
\end{align}
where $ V_1 \simeq 0.027 U_p$. Let us note that the ratio $V_1
/ U_p$ is approximately $14\%$ smaller than the naively estimated
nonorthogonal value $1/32=0.03125$. We also checked that
dimensionality drives the following trend in the ratios $V_1/U_p$:
$0.023$, $0.025$, $0.027$ (considered here), and $0.029$, for a
single rung, the 1D case, a ladder, and the 2D case, so $V_1/U_p$
increases with the increasing number of neighbors.\cite{Fei95}
This follows because in lower dimensions the charge escapes easier
from the orbitals $b$ and $y$ (providing the dominating
contribution), while in the 2D case {\it all} the orbitals suffer
from the orthogonality problem.

%%%%%%%%%%%%%%%%%%%%%%%%%%%%%%%%%%%%%%%%%%%%%%%%%%%%%%%%
%%                       figure 2
%%%%%%%%%%%%%%%%%%%%%%%%%%%%%%%%%%%%%%%%%%%%%%%%%%%%%%%%

\subsection{Interladder repulsion $H_{V_2}$}

Finally, we calculate the {\it interladder} repulsion
between the ZR singlets due to on-site repulsion
$U_p$ in orbitals belonging to two neighboring ladders:
$\langle \psi_{s \alpha}, \bar{\psi}_{r+\frac12, \beta} | \mathcal{H}_2 |
\bar{\psi}_{h+\frac12, \beta }, \psi_{j \alpha} \rangle$
--- a {\it bar} sign over $\psi$ denotes the singlet formed on the
neighboring ladder. Besides, since the neighboring ladder is
misaligned by a lattice constant $1/2$ with respect to the one
considered, we label the ZR singlets on the neighboring ladder by
$j+1/2$ (for the copper-copper lattice constant equal to $1$).
Next, using Eq. (\ref{eq:ZhangRice}) one finds after a somewhat
tedious but straightforward calculation (for more details see
Appendix \ref{app:b}) the following value for the interladder
interaction between the closest sites belonging to the neighboring
ladders (see Fig. \ref{fig:2}),
\begin{equation}
 \langle \psi_{j\alpha},\! \bar{\psi}_{j\pm \frac12,\bar{\alpha}} | \mathcal{H}_2 |
\bar{\psi}_{j\pm \frac12,\bar{\alpha}},\! \psi_{j\alpha } \rangle\!=\!0.136\,
(1\!-\!5\eta/2)\,U_p\,,
\end{equation}
while other (neglected) longer-range repulsive terms are
at least one order of magnitude smaller, cf. Appendix \ref{app:b}.
Thus, the repulsion between holes on the neighboring ladders
reads:
\begin{align}\label{eq:hv2}
H_{V_2}=V_2 \sum_{i\alpha}  \Big( \tilde{n}_{i \alpha}
\tilde{\bar{n}}_{i+\frac12, \bar{\alpha}}
+ \tilde{n}_{i\alpha} \tilde{\bar{n}}_{i-\frac12,
\bar{\alpha}} \Big)\,,
\end{align}
where  $\tilde{n}_{i \alpha}$ operator is related to the ZR
singlets as before and $V_2 \simeq 0.136\,(1-5\eta/2)\,U_p$. We
again neglected all spin-flip terms which are small in comparison
with the ZR binding energy and give zero when `sandwiched' in the
singlet states. Besides, the numerical prefactor (equal to
$0.136$) is here slightly enhanced with respect to the expected
$1/8=0.125$ value (unlike in the intraladder case). This is
because a significant fraction of charge escapes from the $b$ and
$y$ orbitals to the $x$ orbitals due to the orthogonalization
procedure.

\begin{table}[t!] \caption{Adopted values of the parameters of the
charge transfer model (\ref{eq:ct}) from Ref. \onlinecite{Hyb89}
($J_H$ from Ref. \onlinecite{Gra92})
and the calculated values of the derived $t$--$J$--$V$ model
(\ref{eq:tj}) in eV.} \label{tab:2}
\begin{ruledtabular}
\begin{tabular}{c c c c}
\multicolumn{2}{c}{charge transfer model} &
\multicolumn{2}{c}{$t$--$J$--$V$ model} \cr \hline
 $t_{pd}$  & 1.3  &   $t$      &     0.54   \cr
 $\Delta$  & 3.6  &   $J$      &     0.24   \cr
 $U_p$     & 4.0  &   $V_1$    &     0.11   \cr
 $U$       & 10.5 &   $V_2$    &     0.27   \cr
 $J_H$     & 0.8  &            &
\end{tabular}
\end{ruledtabular}
\end{table}

\subsection{Parameters of the effective model}
\label{sec:para}

The calculated parameters of the effective $t$--$J$--$V$ model
(\ref{eq:tj}) derived above are shown in Table \ref{tab:2}. This
calculation is based on the cuprate charge transfer model
parameters from Ref. \onlinecite{Hyb89} (the value of $J_H$ is
taken from Ref. \onlinecite{Gra92}) which might be considered as
the most widely accepted choice of the cuprate parameters, cf.
Refs. \onlinecite{Arr09} and \onlinecite{Fei95}.
Due to the same Cu--O distances in SCCO as in CuO$_2$ planes in
e.g. La$_2$CuO$_4$, we can can adopt these parameters also to the
present case.

Let us note that on the one hand, it should be emphasized that the
{\it interladder\/} coupling $V_2$ is of the order of $J$ for the
realistic parameters\cite{Hyb89,Gra92} and therefore {\it cannot be
neglected\/}. On the other hand, the value of $V_1$ is two and a half
times smaller and therefore we suggest that, if necessary, this
interaction could be skipped in the first-order calculations.

\section{Role of the intersite Coulomb repulsion $V_{pd}$}
\label{sec:4}

One may wonder whether the intersite Coulomb repulsion $V_{pd}$ in
the charge transfer model could alone lead to a significant
repulsion (i.e., of the order of the estimated value of $V_2$)
between the ZR singlets in the neighboring ladders. This term,
which stands for the repulsion between charges situated in the
nearest neighbor copper $3d$ and oxygen $2p$ orbitals, was
neglected in Ref. \onlinecite{Woh07} and in the above analysis,
cf. Eq. (\ref{eq:ct}). In fact, including this term may e.g. lead
to a significant renormalization of the parameters of the 2D
$t$--$J$ model.\cite{Bel94}

%%%%%%%%%%%%%%%%%%%%%%%%%%%%%%%%%%%%%%%%%%%%%%%%%%%%%%%%
%%                       figure 3
%%%%%%%%%%%%%%%%%%%%%%%%%%%%%%%%%%%%%%%%%%%%%%%%%%%%%%%%
\begin{figure*}[t!]
\includegraphics[width=0.75\textwidth]{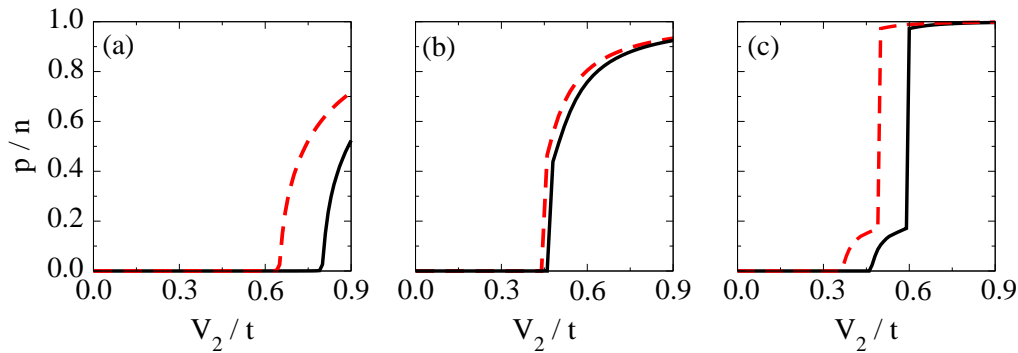}
\caption{(Color online) The self-consistently calculated order
parameter $p$
for SCCO as obtained from the effective $t$--$J$--$V$ model
(\ref{eq:tj})
[see Eq. (\ref{eq:p}) and text for more details] as a function of
the interladder interaction $V_2$ for: (a) filling $n=2/3$ and
period $\lambda=3$, (b) $n=3/4$ and $\lambda=4$, (c) $n=4/5$ and
$\lambda=5$. Parameters: realistic values (see text) $J=0.4t$ and
$V_1=0.2t$ (solid lines), and $J=0$ and $V_1=0$ (dashed lines).}
\label{fig:3}
\end{figure*}

Indeed, one finds that the repulsion between a hole in a symmetric
state $|\phi_{i \sigma}\rangle$ and a copper hole in state $|d_{j
\sigma'}\rangle$ situated on the nearest neighbor sites in two
neighboring ladders is of the order of $
(n_{ix\sigma}+n_{iy\sigma}) n_{dj\sigma'} V_{pd} \sim 0.5 V_{pd}$. Although
typically $V_{pd}$ is smaller than $t_{pd}$, e.g. 
$V_{pd}\sim 1.2$ eV \cite{Hyb89} or even $V_{pd} \le 1$ eV,\cite{Esk89}
this contribution might still be significant and, in principle, 
should not be entirely neglected. 
However, the key observation is that this
term leads to roughly equally large energy cost if: (i) {\it
either\/} the two ZR singlets are situated on the above mentioned
sites $i$ and $j$ and repel each other due to $V_{pd}$, {\it or}
(ii) the two ZR singlets are situated far away from each other on
two ladders and (due to $V_{pd}$) merely feel the repulsion with
the neighboring copper hole on the neighboring ladder. Hence,
including the intersite repulsion $V_{pd}$ increases the total
energy of the system but almost does not contribute to the energy
difference between the two above situations, measured by the value
of the interladder repulsion $V_2$. It is only a small residual
repulsion due to the orthogonalization procedure, see Eq.
(\ref{eq:plaql}), which may change the value of $V_2$ by a small
fraction. This has been also confirmed by the results of Ref.
\onlinecite{Fei95}, where $V_{pd}$ leads indeed to a very small
repulsion between holes in the 2D $t$--$J$ model (ca. $0.05
t_{pd}$).

\section{Numerical results}
\label{sec:5}

Now we shall verify whether the derived $t$--$J$--$V$ model Eq.
(\ref{eq:tj}) supports the CDW states observed in SCCO and to
understand to what extent the interladder term (\ref{eq:hv2})
influences the stability of the CDW state. As a thorough
investigation is beyond the scope of this work and left for future
studies, we solve the model (\ref{eq:tj}) in the simplest possible
way. Thus, we first introduce the Gutzwiller factors,
$g_t=(2-2n)/(2-n)$ and $g_J=4/(2-n)^2$, which renormalize the
kinetic ($g_t$) and interaction ($g_J$) terms; for their
justification see e.g. Refs. \onlinecite{Rac07}. Here $n$ denotes
the average number of $d$ holes per site in the effective model
(\ref{eq:tj}), i.e., $n=\sum_\sigma \langle
\tilde{n}_{i\alpha\sigma}\rangle$. Second, we use the mean-field
approximation for the interaction terms. Next, we diagonalize the
effective one-particle Hamiltonian introducing the classical
fields
\begin{equation}\label{eq:p}
\langle d^{\dag}_{i\alpha\sigma}d^{}_{i\alpha\sigma}\rangle =\left\{
\begin{array}{cc}
n-p & {\rm for}\  i/\lambda \in \mathbb{Z} \\
n+\frac{1}{\lambda-1} p & {\rm for }\ i/\lambda \notin\mathbb{Z}
\end{array} \right. \,,
\end{equation}
where $p\le n$ is the CDW order parameter and $\lambda$ is the CDW
period. Furthermore,
$\langle\bar{d}^{\dag}_{i-\frac12,\alpha\sigma}\bar{d}_{i-\frac12,\alpha
\sigma}\rangle$ are defined as in Eq. (\ref{eq:p}) but with $i/\lambda$
replaced by $(i+1)/\lambda$ [$(i+2)/\lambda$ for $\lambda=5$] ---
this assumption minimizes the classical energy cost of the interladder
repulsion $V_2$.

In Fig. \ref{fig:3} we show the CDW order parameter $p$ calculated
self-consistently as a function of the interladder interaction
$V_2$ for the three experimentally interesting doping
levels:\cite{Woh07,Rus07} $n=2/3$ which corresponds to
$n_h=2-n=4/3$ holes per copper site in charge transfer model
(\ref{eq:ct}), $n=3/4$ corresponding to $n_h=5/4$, and $n=4/5$
corresponding to $n_h=6/5$. The results demonstrate that the
interladder interaction plays indeed a crucial role in the
stability of the CDW while the intraladder one is rather
unimportant.

We have found that the CDW state with period $\lambda=3$
($\lambda=5$) is stable for $n=2/3$ ($n=4/5$) for the rather
realistic values of the parameters $J=0.4t$, $V_1=0.2t$ and
$V_2=0.5t-0.9t$. Please note, that: (i) we adopted here a somewhat
smaller value of $J=0.4t$ which is closer to a typical value for
cuprates,\cite{Poi10} and (ii) values of $V_2$ exceeding $0.5$t
can be obtained using for example the set of parameters suggested
in Ref. \onlinecite{Esk89}. This finding explains well the
experimental results of Ref. \onlinecite{Rus06}. Besides, the CDW
ordered state is also stable for period $\lambda=4$ which was not
observed.\cite{Rus06} We expect that the stability of the CDW
state with this period is a shortcoming of the above simplified
solution which does not capture well the frustration between two
possible CDW patterns in the neighboring ladders which occurs for
period $\lambda=4$.\cite{Woh07}

\section{Conclusions}
\label{sec:6}

In summary, we derived the $t$--$J$ model which describes the low
energy physics of Cu$_2$O$_5$ coupled ladders. Apart from the
`standard' superexchange $\propto J$ and kinetic energy terms
$\propto t$, the model contains also the repulsion between the
nearest neighbor holes in a ladder $\propto V_1$ and in the two
neighboring ladders $\propto V_2$, and hence is also referred to
as a $t$--$J$--$V$ model [see Eq. (\ref{eq:tj})]. We showed that the latter $V_2$
term is roughly two and a half times larger than $V_1$ and (contrary
to $V_1$) cannot be skipped --- in fact it is crucial to explain the
onset of the odd period CDW state in SCCO. We emphasize that this
particular extra term is restricted to the copper oxides in which
oxygen is coordinated by three copper ions in the same plane.
Therefore, it is not present in the CuO$_2$ planes,\cite{Ole87}
or in Cu--O chain\cite{Ole91} in copper oxides, but
(apart from the discussed SCCO case) could become
relevant for the coupled chains of SrCuO$_2$ (provided they are
hole-doped). Furthermore, it is both a many-body term {\it and\/}
of the order of $J$, contrary to various corrections to a 1D or 2D
$t$--$J$ model.\cite{Fei95,Mai05} Besides, we also
verified that the Coulomb intersite interacton $V_{pd}$ alone (not
included in the presented derivation) cannot lead to a significant
interladder repulsion $V_2$.

The simple mean-field solutions of the $t$--$J$--$V$ model derived
here provides evidence in favor of the experimental observations
of the onset of the odd CDW state in the ladder planes of the
SCCO. This is further supported by the recent density matrix 
renormalization group calculations
\cite{Poi10} where a ladder $t$--$J$ model with the interladder
coupling $V_2$ (denoted as $V_{\perp}$ in Ref. \onlinecite{Poi10})
was studied: also there the CDW with odd period is stabilized due
to the presence of the interladder interaction $V_2$.

Finally, let us note that the other hole-doped ladder compound
La$_{1-x}$Sr$_x$CuO$_{2.5}$ is also characterized by a large (but
different than the one discussed here) interladder
coupling.\cite{Mat96} Therefore, we argue that it is currently a
challenge for the condensed matter community to search for a
hole-doped ladder compound which could indeed be modelled by the
ladder $t$--$J$ Hamiltonian [given by Eqs. (\ref{eq:hj}) and
(\ref{eq:ht}), i.e., without additional intersite repulsion
terms].

\acknowledgments
We thank Alexander Chernyshev for insightful discussions
and Maria Daghofer for the critical reading of the manuscript.
We acknowledge financial support by the Foundation for Polish
Science (FNP) and the Polish Ministry of Science and Education
under Project No.~N202 068 32/1481.
K. W. thanks University of British Columbia for the kind hospitality.

\appendix
\section{Derivation of the intraladder repulsion term $\propto V_1$}
\label{app:a}

Here we show how to calculate the repulsion between orthogonalized
ZR singlets within the ladder due to the on-site interaction $U_p$
in $\mathcal{H}_1$. Thus, one needs to determine the following
matrix elements:
\begin{equation}\label{eq:matrix}
\langle \psi_{s\alpha}, \psi_{r\alpha} | \mathcal{H}_1 |
\psi_{h\alpha}, \psi_{j\alpha} \rangle, \quad
\langle \psi_{s\alpha}, \psi_{r\bar{\alpha}} | \mathcal{H}_1 |
\psi_{h\bar{\alpha}}, \psi_{j{\alpha}} \rangle.
\end{equation}
Let us note that the mixed terms such as for example $\langle
\psi_{s R}, \psi_{r L} | \mathcal{H}_1 | \psi_{h L}, \psi_{jL}
\rangle$ vanish in the ZR singlet basis -- they could {\it a
priori\/} lead to the destruction of the ZR singlets, but
fortunately they are much smaller than the respective binding
energy.

{\it Intraladder repulsion along the leg.---}
First, we calculate the matrix elements of $\mathcal{H}_1$ between the
orthogonal plaquette states Eq. (\ref{eq:plaql}) along the leg:
\begin{align}
&\langle \phi_{s\alpha\sigma}, \phi_{r\alpha\bar{\sigma}} | \mathcal{H}_1 |
\phi_{h\alpha\bar{\sigma}}, \phi_{j\alpha {\sigma}} \rangle = \nonumber \\
&\frac{1}{16} U_p \frac{1}{N^3}\sum_{k q f} e^{ik(h-r)}e^{iq(j-s)}e^{if(r-s)} \nonumber \\
&\times\Big\{ \frac{1}{16}\Big(\alpha_k\alpha_q + \beta_k\beta_q
                        - \alpha_k\beta_q - \beta_k\alpha_q\Big) \nonumber \\
&\times\Big(\alpha_{q-f}\alpha_{k+f} + \beta_{q-f}\beta_{k+f}
          - \alpha_{q-f}\beta_{k+f} - \beta_{q-f}\alpha_{k+f}\Big) \nonumber \\
&+ \Big(\sin\frac{k}{2} \sin\frac{q}{2} \sin\frac{q-f}{2} \sin\frac{k+f}{2}
+ \frac{1}{16} \Big) \nonumber \\
&\times\Big(\alpha_k\alpha_q \alpha_{q-f}\alpha_{k+f}
+ \beta_k\beta_q \beta_{q-f}\beta_{k+f} \Big) \Big\},
\end{align}
and
\begin{eqnarray}
\label{eq:negative}
&&\langle \phi_{s\alpha\sigma}, \phi_{r\alpha\bar{\sigma}} | \mathcal{H}_1 |
\phi_{h\alpha{\sigma}}, \phi_{j\alpha \bar{\sigma}} \rangle \nonumber \\
&=& -\langle \phi_{s\alpha\sigma}, \phi_{r\alpha\bar{\sigma}} | \mathcal{H}_1 |
\phi_{h\alpha\bar{\sigma}}, \phi_{j\alpha {\sigma}} \rangle,
\end{eqnarray}
while the same spin elements are zero,
\begin{equation}
\langle \phi_{s\alpha\sigma}, \phi_{r\alpha{\sigma}} | \mathcal{H}_1 |
\phi_{h\alpha{\sigma}}, \phi_{j\alpha {\sigma}} \rangle = 0.
\end{equation}
One can evaluate numerically the above expressions. It occurs that
the largest positive element is the nearest neighbor interaction
\begin{align} \label{eq:phileg}
 \langle \phi_{j\alpha\sigma}, \phi_{j+1,\alpha\bar{\sigma}} | \mathcal{H}_1 |
\phi_{j+1,\alpha\bar{\sigma}}, \phi_{j\alpha {\sigma}}
\rangle=0.0544\,U_p,
\end{align}
while following Eq. (\ref{eq:negative}) the absolute value of the
largest negative element, which corresponds to spin-flip nearest
neighbor interaction, is the same. Furthermore, the second largest
element is the next nearest neighbor interaction and is over 20
times smaller, which means that it can be safely neglected.

Second, we calculate the matrix elements of $\mathcal{H}_1$
between the nearest neighbor ZR singlets, defined by
Eq. (\ref{eq:ZhangRice}). This introduces a factor
$1/2$ to the above estimations of the repulsion between orthogonal
plaquette states: It is because there is a $50\%$ probability to
have opposite spins on a particular shared oxygen site occupied by
two holes from two different ZR singlets. Note that the
spin-flip-plaquette terms do not give any contribution to the
repulsion between ZR singlets, although they could in principle
destabilize the ZR states themselves. Fortunately, this is not
possible since the binding energy of the ZR singlets is much
larger. Thus altogether, we obtain for the repulsion along the
same leg
\begin{align} \label{eq:psileg}
 \langle \psi_{j\alpha}, \psi_{j+1,\alpha} | \mathcal{H}_1 |
\psi_{j+1,\alpha}, \psi_{j\alpha } \rangle=0.0272\,U_p.
\end{align}

{\it Intraladder repulsion along the rung.---} Following a similar
scheme, one can calculate the repulsion between ZR singlets on
different legs. One obtains the following matrix elements of
$\mathcal{H}_1$ between the orthogonal plaquette states Eq.
(\ref{eq:plaql}) on different legs
\begin{align}
&\langle \phi_{s\alpha\sigma}, \phi_{r\bar{\alpha}\bar{\sigma}} | \mathcal{H}_1 |
 \phi_{h\bar{\alpha}\bar{\sigma}}, \phi_{j\alpha {\sigma}} \rangle = \nonumber \\
&\frac{1}{16} U_p \frac{1}{N^3}\sum_{k q f} e^{ik(h-r)}e^{iq(j-s)}e^{if(r-s)} \nonumber \\
&\times\Big\{ \frac{1}{16}\Big(\alpha_k\beta_q + \beta_k\alpha_q
             -\alpha_k\alpha_q -\beta_k\beta_q \Big) \nonumber \\
&\times\Big(\alpha_{q-f}\beta_{k+f} + \beta_{q-f}\alpha_{k+f}
-\alpha_{q-f}\alpha_{k+f} - \beta_{q-f}\beta_{k+f}  \Big) \nonumber \\
&+ \Big(\sin\frac{k}{2} \sin\frac{q}{2} \sin\frac{q-f}{2} \sin\frac{k+f}{2}
+ \frac{1}{16} \Big) \nonumber \\
&\times\Big(\alpha_k\beta_q \alpha_{q-f}\beta_{k+f}
+ \beta_k\alpha_q \beta_{q-f}\alpha_{k+f} \Big) \Big\},
\end{align}
and
\begin{eqnarray}
&&\langle \phi_{s\alpha\sigma}, \phi_{r\bar{\alpha}\bar{\sigma}} | \mathcal{H}_1 |
\phi_{h\bar{\alpha}{\sigma}}, \phi_{j\alpha \bar{\sigma}} \rangle  \nonumber \\
&=& -\langle \phi_{s\alpha\sigma}, \phi_{r\bar{\alpha}\bar{\sigma}} | \mathcal{H}_1 |
\phi_{h\bar{\alpha}\bar{\sigma}}, \phi_{j\alpha {\sigma}} \rangle,
\end{eqnarray}
and
\begin{equation}\label{eq:negative2}
\langle \phi_{s\alpha\sigma}, \phi_{r\bar{\alpha}{\sigma}} | \mathcal{H}_1 |
\phi_{h\bar{\alpha}{\sigma}}, \phi_{j\alpha {\sigma}} \rangle = 0.
\end{equation}
Evaluating numerically the above expressions one obtains that the
largest element is the nearest neighbor repulsion --- this time
between the orthogonal plaquette states on the same rung:
\begin{align}
 \langle \phi_{j\alpha\sigma}, \phi_{j,\bar{\alpha}\bar{\sigma}} | \mathcal{H}_1 |
\phi_{j,\bar{\alpha}\bar{\sigma}}, \phi_{j\alpha {\sigma}}
\rangle=0.0529\,U_p,
\end{align}
while the second largest element (the next nearest neighbor
interaction) is over 15 times smaller and can be
neglected.

Finally, following the same steps as those leading from (\ref{eq:phileg}) 
to (\ref{eq:psileg}), we obtain the repulsion between
the nearest neighbor ZR singlets [defined by Eq. (\ref{eq:ZhangRice})]
along the same rung which is twice reduced:
\begin{align}
 \langle \psi_{j\alpha}, \psi_{j,\bar{\alpha}} | \mathcal{H}_1 |
\psi_{j,\bar{\alpha}}, \psi_{j\alpha } \rangle=0.0265\,U_p.
\end{align}

\section{Derivation of the interladder repulsion term $\propto V_2$}
\label{app:b}

Here the task is to calculate the repulsion between two ZR
singlets centered at the neighboring copper positions of two
ladders (and thus sharing the same oxygen sites but {\it not} the
$p$ orbitals, see Fig. \ref{fig:2}) due to the on-site repulsion
on oxygen sites. However, again we will calculate the repulsion
between arbitrarily located ZR singlets and only then we will show
which elements are negligible. Note that the plaquette states on
two ladders are orthogonal to each other although they still have
to be orthogonalized for the same ladder (as in Appendix \ref{app:a}).
Explicitly one needs to calculate the following matrix elements:
\begin{equation}
\langle \psi_{s \alpha}, \bar{\psi}_{r+\frac12, \bar{\alpha}} | \mathcal{H}_2 |
\bar{\psi}_{h+\frac12, \bar{\alpha} }, \psi_{j \alpha} \rangle,
\end{equation}
and
\begin{equation}
\langle \psi_{s \alpha}, \bar{\psi}_{r+\frac12, {\alpha}} | \mathcal{H}_2 |
\bar{\psi}_{h+\frac12, {\alpha} }, \psi_{j \alpha} \rangle.
\end{equation}

{\it Interladder repulsion between plaquettes with the same
spin.---} We calculate the matrix elements of
$\mathcal{H}_2$ between the orthogonal plaquette states Eq.
(\ref{eq:plaql}) with the same spin but situated on different
legs:
\begin{align}
&\langle \phi_{r \alpha \sigma}, \bar{\phi}_{s+\frac12, \bar{\alpha} \sigma} |
\mathcal{H}_2  | \bar{\phi}_{h+\frac12, \bar{\alpha} \sigma}, \phi_{j\alpha \sigma}
\rangle= \nonumber \\
&\frac{1}{16}(1-3\eta)U_p\frac{1}{N^3}\sum_{k q f}
\alpha_{k} \alpha_{q} \alpha_{q-f} \alpha_{k+f} \nonumber \\
&\times\Big\{ \frac{1}{4} \sin q  \sin(q-f) +\frac{1}{4} \sin k \sin(k+f)\Big\} \nonumber \\
&\times e^{ik(h-r)}e^{iq(j-s)}e^{if(r-s-\frac12)},
\end{align}
while for the same legs we obtain
\begin{align}
&\langle \phi_{r \alpha \sigma}, \bar{\phi}_{s+\frac12, {\alpha} \sigma} | \mathcal{H}_2  |
\bar{\phi}_{h+\frac12, {\alpha} \sigma}, \phi_{j\alpha \sigma}\rangle = \nonumber \\
&\frac{1}{16}(1-3\eta)U_p\frac{1}{N^3}\sum_{k q f}
\alpha_{k} \beta_{q} \alpha_{q-f} \beta_{k+f} \nonumber \\
&\times \Big\{ \frac{1}{4} \sin q  \sin(q-f)
+\frac{1}{4} \sin k \sin(k+f)\Big\} \nonumber \\
&\times e^{ik(h-r)}e^{iq(j-s)}e^{if(r-s-\frac12)}.
\end{align}
As it might have been expected, it occurs that the biggest term is the
repulsion between orthogonal plaquette states with the same spin situated
on the closest possible sites in the neighboring ladders (see Fig. \ref{fig:2}):
\begin{align}
\label{eq:phiphisame}
\langle \phi_{j \alpha \sigma}, \bar{\phi}_{j \pm \frac12,
\bar{\alpha} \sigma} | \mathcal{H}_2  | \bar{\phi}_{j \pm \frac12,
\bar{\alpha} \sigma}, \phi_{j \alpha \sigma}
\rangle=0.1355\,(1-3\eta)U_p,
\end{align}
and all other terms are of the order of $10^{-3}(1-3\eta)U_p$ and can be neglected.

{\it Interladder repulsion between plaquettes with opposite
spin.---} A very similar calculation as above, but
performed for the orthogonal plaquette states Eq. (\ref{eq:plaql})
with opposite spins leads to the repulsion between orthogonal
plaquette states with opposite spins and situated on the closest
possible sites in the neighboring ladders:
\begin{align}
\label{eq:phiphiopposite}
\langle \phi_{j \alpha \sigma}, \bar{\phi}_{j \pm \frac12,
\bar{\alpha} \bar{\sigma}} | \mathcal{H}_2  | \bar{\phi}_{j \pm
\frac12, \bar{\alpha} \bar{\sigma}}, \phi_{j \alpha \sigma}
\rangle=0.1355\,(1-2\eta) U_p,
\end{align}
while again all other longer-range repulsive terms can be neglected.

Finally, combining Eqs. (\ref{eq:phiphisame})-(\ref{eq:phiphiopposite})
with the definition of the ZR singlet (\ref{eq:ZhangRice}) we obtain the
value of the repulsion between the two ZR singlets 
(cf. similar discussion in Appendix \ref{app:a}) on the closest
possible sites in the neighboring ladders to be
\begin{equation}
 \langle \psi_{j\alpha},\! \bar{\psi}_{j\pm \frac12,\bar{\alpha}} | \mathcal{H}_2 |
\bar{\psi}_{j\pm \frac12,\bar{\alpha}},\! \psi_{j\alpha } \rangle\!=\!0.1355\,
(1\!-\!5\eta/2)\,U_p\,.
\end{equation}

%%%%%%%%%%%%%%%%%%%%%%%%%%%%%%%%%%%%%%%%%%%%%%%%%%%%%%%%%%%%%%%%%%%
%%                             REFERENCES
%%%%%%%%%%%%%%%%%%%%%%%%%%%%%%%%%%%%%%%%%%%%%%%%%%%%%%%%%%%%%%%%%%%


\begin{thebibliography}{00}

\bibitem{Cha78} K. A. Chao, J. Spa\l{}ek, and A. M. Ole\'s,
                   J. Phys. C \textbf{10}, L271 (1977);
                   \prb \textbf{18}, 3453 (1978).

\bibitem{Zha88} F. C. Zhang and T. M. Rice,
                   \prb \textbf{37}, 3759 (1988).

\bibitem{Dag94} E. Dagotto,
                   \rmp \textbf{66}, 763, (1994).

\bibitem{Mai05} T. A. Maier, M. Jarrell, T. C. Schulthess,
                   P. R. C. Kent, and J. B. White,
                   \prl \textbf{95}, 237001 (2005);
                M. Ogata and H. Fukuyama,
                   Rep. Prog. Phys. \textbf{71}, 036501 (2008);
                L. Spanu, M. Lugas, F. Becca, and S. Sorella,
                   \prb \textbf{77}, 024510 (2008).

\bibitem{Dag96} E. Dagotto and T. M. Rice,
                   Science \textbf{271}, 618 (1996).

\bibitem{Rus06} A. Rusydi, P. Abbamonte, H. Eisaki, Y. Fujimaki,
                   G. Blumberg, S. Uchida, and G. A. Sawatzky,
                   \prl \textbf{97}, 016403 (2006).

\bibitem{Dag92} E. Dagotto, J. Riera, and D. J. Scalapino,
                   \prb \textbf{45}, 5744 (1992).

\bibitem{Ueh96} M. Uehara, T. Nagata, J. Akimitsu, H. Takahashi,
                   N. M\^ori, and K. Kinoshita,
                   J. Phys. Soc. Jpn. \textbf{65}, 2764 (1996).

\bibitem{Woh07} K. Wohlfeld, A. M. Ole\'s, and G. A. Sawatzky,
                   \prb \textbf{75}, 180501 (2007).

\bibitem{Whi02} S. R. White, I. Affleck, and D. J. Scalapino,
                   \prb \textbf{65}, 165122 (2002).

\bibitem{Rou07} G. Roux, E. Orignac, S. R. White, and D. Poilblanc,
                   \prb \textbf{76}, 195105 (2007).

\bibitem{Ole87} V. J. Emery,
                   \prl \textbf{58}, 2794 (1987);
                A. M. Ole\'s, J. Zaanen, P. Fulde,
                   Physica B\&C \textbf{148}, 260 (1987);
                C. M. Varma, S. Schmitt Rink, and E. Abrahams,
                   Solid State Commun. \textbf{62}, 681 (1987);
                J. Dutka and A.~M. Ole\'s,
                   \prb \textbf{42}, 105 (1990).

\bibitem{Ole91} A. M. Ole\'s and W. Grzelka,
                   \prb \textbf{44}, 9531 (1991).

\bibitem{Gra92} J. B. Grant and A. K. McMahan,
                   \prb \textbf{46}, 8440 (1992).

\bibitem{Mue98} T. F. A. M\"uller, V. Anisimov, T. M. Rice, I. Dasgupta,
                   and T. Saha-Dasgupta,
                   \prb \textbf{57}, R12655 (1998).

\bibitem{Woh07b} K. Wohlfeld,
                   AIP Conf. Proc. \textbf{918}, 337 (2007).

\bibitem{Hyb89} M. S. Hybertsen, M. Schl\"uter, and N. E. Christensen,
                   \prb \textbf{39}, 9028 (1989).

\bibitem{Zaa88} J. Zaanen and A. M. Ole\'s,
                   \prb \textbf{37}, 9423 (1988).

\bibitem{Kan59} J. Kanamori,
                   J. Phys. Chem. Solids \textbf{10}, 87 (1959);
                   J. B. Goodenough, {\it Magnetism and the Chemical Bond\/}
                   (Interscience, Wiley, 1963).

\bibitem{symm}  The term `symmetric' refers here to the $d_{x^2-y^2}$ symmetry.

\bibitem{Arr09} E. Arrigoni, M. Aichhorn, M. Daghofer, and W. Hanke,
                   New J. Phys. \textbf{11}, 055066 (2009).

\bibitem{Fei95} L. F. Feiner, J. H. Jefferson, and R. Raimondi,
                   \prb \textbf{51}, 12797 (1995);
                        \textbf{53},  8751 (1996);
                R. Raimondi, J. H. Jefferson, and L. F. Feiner,
                   {\it ibid.\/} \textbf{53}, 8774 (1996).

\bibitem{Bel94} V. I. Belinicher and A. L. Chernyshev,
                   \prb \textbf{49}, 9746 (1994).

\bibitem{Esk89} H. Eskes, G. A. Sawatzky, and L. F. Feiner,
                   Physica C \textbf{160}, 424 (1989).

\bibitem{Rac07} F. C. Zhang,
                   \prl \textbf{90}, 207002 (2003);
                J. Y. Gan, Y. Chen, Z. B. Su, and F. C. Zhang,
                   {\it ibid.\/} \textbf{94}, 067005 (2005);
                M. Raczkowski, M. Capello, D. Poilblanc, R. Fr\'esard,
                   and A. M. Ole\'s,
                   \prb \textbf{76}, 140505 (2007).

\bibitem{Rus07} A. Rusydi, M. Berciu, P. Abbamonte, S. Smadici,
                   H. Eisaki, Y. Fujimaki, S. Uchida, M. R\"ubhausen,
                   and G. A. Sawatzky,
                   \prb \textbf{75}, 104510 (2007).

\bibitem{Poi10} J. Almeida, G. Roux, and D. Poilblanc,
                   arXiv:1002.4367v1 (unpublished).

\bibitem{Mat96} In La$_{1-x}$Sr$_x$CuO$_{2.5}$ the interladder coupling
                   even leads to the antiferromagnetic order for $x=0$,
                   see: S. Matsumoto, Y. Kitaoka, K. Ishida, K. Asayama,
                   Z. Hiroi, N. Kobayashi, and M. Takano,
                   \prb \textbf{53}, R11942 (1996).


\end{thebibliography}
\end{document}